\documentstyle[12pt]{article}
\def\IR{{\bf R}}
\catcode`\@=11

\long\def\@makefntext#1{\parindent 0cm\noindent
\hbox to 1em{\hss$^{\@thefnmark}$}#1}
\begin{document}
\begin{titlepage}
\vspace{.5in}
\begin{flushright}
UCD-94-18\\
February 1994\\
gr-qc/9405043\\
\end{flushright}
\vspace{.5in}
\begin{center}
{\Large\bf
Time in (2+1)-Dimensional Quantum Gravity}\\
\vspace{.4in}
{S.~C{\sc arlip}\footnote{\it email: carlip@dirac.ucdavis.edu}\\
       {\small\it Department of Physics}\\
       {\small\it University of California}\\
       {\small\it Davis, CA 95616}\\{\small\it USA}}
\end{center}

\vspace{.5in}
\begin{center}
{\large\bf Abstract}
\end{center}
\begin{center}
\begin{minipage}{5in}
{\small
General relativity in three spacetime dimensions is used
to explore three approaches to the ``problem of time'' in quantum gravity:
the internal Schr\"odinger approach with mean extrinsic curvature as a time
variable, the Wheeler-DeWitt equation, and covariant canonical quantization
with ``evolving constants of motion.''  (To appear in {\em Proc.\
of the Lanczos Centenary Conference}, Raleigh, NC, December 1993.)
}
\end{minipage}
\end{center}
\end{titlepage}
\addtocounter{footnote}{-1}

As Karel Kucha{\v r} has explained elsewhere in these Proceedings, the
nature of time in quantum gravity is at best obscure.  Straightforward
attempts at quantization lead to such absurdities as vanishing
Hamiltonians and undefined inner products.  The problems are not merely
technical, but reflect an underlying conceptual issue: time translations
in general relativity are ``gauge symmetries,'' and do not
have an obvious physical meaning.

It may be that this problem cannot be resolved without a full-fledged
quantum theory of gravity.  But our confusion about the nature of time is
itself an obstacle to the construction of such a theory.  It is therefore
useful to look for simpler models to explore possible approaches to the
role of time in quantum gravity.

General relativity in 2+1 dimensions is one such model.  Like realistic
(3+1)-dimensional gravity, it is a generally covariant theory of spacetime
geometry.  But  (2+1)-dimensional general relativity has only finitely many
physical degrees of freedom, reducing the problem of quantization from
quantum field theory to more elementary quantum mechanics.

\section{Classical Theory}

Let us start by investigating the nature of time in classical
(2+1)-dimensional relativity.  The basic simplification of 2+1 dimensions
is that the full curvature tensor is determined by the Ricci tensor;
in particular, any empty space solution of the Einstein equations is
{\em flat}.  For a spacetime with trivial topology, the flat metric is
unique up to diffeomorphisms, and no dynamics is left.  For a topologically
nontrivial spacetime, however, a finite number of global geometric degrees
of freedom remain.

We can describe these degrees of freedom in two ways, in which time is
treated quite differently:

{\bf 1.\ }The vanishing of $R_{abcd}$ means that any point in a spacetime
$M$ is contained in a coordinate patch isometric to flat Minkowski space
with the standard metric $\eta_{ab}$.  The transition function between two
such patches must be an isometry of $\eta_{ab}$, i.e., an element of the
Poincar\'e group ISO($2,1$).  One such transition function is required for
each noncontractible curve in $M$; in fact, these functions determine a
group homomorphism
\begin{equation} \label{a2}
\rho: \pi_1(M)\rightarrow \hbox{ISO($2,1$)} ,
\end{equation}
the holonomy, unique up to conjugation by an arbitrary fixed element
of ISO($2,1$) \cite{CarB}.

Such a construction, in which a manifold is built by gluing together
geometric patches with isometries, is known to mathematicians as a
geometric structure, and much can be learned from the mathematical
literature \cite{Thurs,Mess}.  In particular, suppose
$M\approx\IR\!\times\!\Sigma$, where $\Sigma$ is a closed surface of
genus $g\!>\!0$.  Then with appropriate assumptions about causal
structure, a homomorphism (\ref{a2}) determines a unique flat
spacetime.  Moreover, the space of ISO($2,1$) holonomies has the
structure of a cotangent bundle, whose $(6g-6)$-dimensional base space
consists of the SO($2,1$) projections of the homomorphisms (\ref{a2}).
The holonomies thus provide an invariant and time-independent
``topological'' description of the space of solutions of the field
equations, essentially equivalent to that suggested by Witten
\cite{Witten} in the Chern-Simons approach to (2+1)-dimensional gravity.

{\bf 2.\ }A very different approach to solving the field
equations starts with standard metric variables in the ADM formalism.  The
spatial metric on $\Sigma$ may be decomposed as
\begin{equation} \label{a3}
g_{ij} = e^{2\lambda}f^*\bar g_{ij}(\tau^\alpha)
\end{equation}
where $f$ is a spatial diffeomorphism, $e^{2\lambda}$ is an arbitrary
conformal factor, and the $\bar g_{ij}$ are a $(6g-6)$-dimensional family
of ``standard'' metrics of constant negative curvature.  A similar
decomposition exists for the canonical momentum $\pi^{ij}$.  If we
choose York's time slicing, foliating $M$ by surfaces of constant
mean extrinsic curvature $\hbox{Tr}K\!=\!T$, the standard constraints
of general relativity completely determine all of the variables except the
moduli $\tau^\alpha$ and their conjugate momenta $p_\alpha$ \cite{Mon}.
The reduced phase space action becomes
\begin{equation} \label{a4}
S = \int dT\left( p_\alpha\dot\tau^\alpha - H(p,\tau,T) \right),
\end{equation}
with a calculable effective Hamiltonian $H$.

Although it is far from obvious, this metric picture is equivalent to
the description in terms of geometric structures.  Given a set of
holonomies $\rho$, one may construct the spacetime $M$ with these holonomies,
foliate it by surfaces of constant mean curvature, and compute the induced
metric on the slice $\hbox{Tr}K\!=\!T$; the resulting moduli $\tau^\alpha$
then obey the equations of motion coming from the action (\ref{a4}).
Conversely, starting with a set of moduli and momenta,
one may compute the corresponding holonomies --- via the Chern-Simons
formalism, for example --- and show that they give the correct description
of the transition functions.

\section{Quantum Gravity and Time}

Let us now turn to the quantum theory.  The simplest quantization, a
version of the ``internal Schr\"odinger approach,'' starts with the
reduced phase space action (\ref{a4}).  This is the action for an ordinary
finite-dimensional quantum mechanical system, and in principle its
quantization is straightforward: we simply treat $\tau^\alpha$ and
$p_\alpha$ as operators on a Hilbert space of square-integrable functions
of the $\tau^\alpha$, with
\begin{equation} \label{b1}
[\hat\tau^\alpha, \hat p_\beta] = i\hbar\delta^\alpha_\beta .
\end{equation}
There are, to be sure, some practical problems --- the Hamiltonian $H$ is
a nonpolynomial function of the positions and momenta, often known only
implicitly, and there are severe operator ordering ambiguities.  More
interesting is a basic conceptual issue.  In this approach to quantization,
the choice of time slicing is made classically, and only the reduced phase
space variables $\tau$ and $p$ are subject to quantum fluctuations.  It is
not at all clear that different choices of time slicing will lead
to equivalent theories.  But (2+1)-dimensional gravity is simple enough
that an answer no longer seems out of reach.

A second approach to quantization starts with the Wheeler-DeWitt equation.
Rather than solving the Hamiltonian constraint classically, we now impose
it as as an operator restriction on physical states.  In effect, we are
permitting quantum fluctuations of the full spatial metric $g_{ij}$ and
its conjugate momentum, not just the $\tau^\alpha$ and $p_\alpha$.

The resulting expression is surprisingly complex --- in particular, it
contains nonlocal terms coming from the momentum constraints
--- and it is hard to find exact solutions.  Moreover, the proper
choice of an inner product on the space of solutions is not obvious.
It may be shown, however, that at least the simplest operator orderings
and inner products give results that are {\em not} equivalent to those of
reduced phase space quantization \cite{WdW}.

Finally, let us return to the idea of classical spacetimes as geometric
structures, and use this picture as a starting point for covariant
canonical quantization \cite{Ash,Crn}, or ``quantizing the space of
classical solutions.''  Recall that the space of holonomies is a cotangent
bundle, with a base space consisting of the SO(2,1) holonomies.  The
quantization of such a bundle is straightforward: wave functions are
sections of a line bundle over the base space, and cotangent vectors act
as Lie derivatives.  But it is not hard to see that the Hamiltonian in
this picture is {\em zero}, giving us a ``frozen time'' description of
the quantum theory.

In retrospect, this should not have been a surprise: our classical starting
point was also time-independent, since the holonomies describe an entire
spacetime at once.  To see even the classical dynamics, we must pick a
time slicing, label slices by a parameter $T$, and compute the moduli
$\tau(\rho,T)$ as functions of the holonomies $\rho$ and the ``time'' $T$.
Different slicings give different moduli, but this reflects the fact that
genuinely inequivalent geometric questions are being asked.

The implications for quantum dynamics are now clear.  We may interpret
frozen time quantization as a Heisenberg picture, in which states are
indeed time-independent.  To construct time-dependent operators, we choose
a slicing, solve for the classical moduli and momenta, and rewrite them as
functions of operators $\hat\rho$, giving what Rovelli calls ``evolving
constants of motion'' \cite{Rov}.  Eigenfunctions of the moduli
$\hat\tau(\hat\rho,T)$ are then ordinary ``time''-dependent Schr\"odinger
picture wave functions appropriate for the chosen time slicing.

For the case of a spacetime with topology $\IR\!\times\!T^2$, this program
has been completed \cite{Car2}.  Interestingly, the result is not
quite equivalent to reduced phase space quantization.  Clearly, much
remains to be learned from this model.

\vspace{1.5ex}
\begin{flushleft}
\large\bf Acknowledgements
\end{flushleft}

This research was supported in part by National Science Foundation Young
Investigator grant PHY-93-57203 and Department of Energy grant
DE-FG03-91ER40674.

\end{document}